\documentclass[a4paper]{jpconf}
\usepackage{graphicx}
\usepackage{hyperref}
\usepackage{amsmath}
\begin{document}
\title{Chemical freeze-out conditions from net-kaon fluctuations at RHIC}

\author{J M Stafford}

\address{Department of Physics, University of Houston, Houston, TX 77204, USA}

\ead{jmstafford@uh.edu}

\begin{abstract}
We compare the mean-over-variance ratio of the net-kaon distribution calculated within a state-of-the-art hadron resonance gas model to the latest experimental data from the Beam Energy Scan at RHIC by the STAR collaboration. Our analysis indicates that it is not possible to reproduce the experimental results using the freeze-out parameters from the existing combined fit of net-proton and net-electric charge mean-over-variance. The strange mesons need about 10-15 MeV higher temperatures than the light hadrons at the highest collision energies. In view of the recent lambda fluctuation measurements, we predict the net-lambda variance-over-mean at the light and strange chemical freeze-out parameters. We observe that the lambda fluctuations are sensitive to the difference in the freeze-out temperatures established in this analysis. Our results have implications for other phenomenological models in the field of relativistic heavy-ion collisions.

\end{abstract}

\section{Introduction}
Relativistic heavy-ion collisions performed at particle accelerator facilities such as the Large Hadron Collider (LHC) at CERN and the Relativistic Heavy Ion Collider (RHIC) at Brookhaven National Laboratory re-create an early state of the universe that existed microseconds after the Big Bang. This state of matter from the primordial universe is the Quark-Gluon Plasma (QGP), so-called because it is the high temperature and density form of matter in which quarks and gluons are deconfined. The aforementioned accelerators allow for the study of the transition between the QGP and the protons and neutrons that are the building blocks for nuclei which permeate the cosmos. The characterization of this phase transition of strongly-interacting, or QCD, matter has received much attention in the past years from both the theoretical and experimental communities \cite{Borsanyi2013, Asakawa2000, Abelev2009, Abelev2013}. The transformation from ordinary matter to the QGP can be qualitatively described by the different stages of a heavy-ion collision (HIC) \cite{Busza2018,Ratti2018}. When the highly Lorentz-contracted nuclei collide, the energy density of the system is large enough to have deconfined quarks and gluons as in the QGP. Due to the near-perfect fluidity of the QGP, the system can be treated hydrodynamically at this stage. Upon cooling and expansion, the energy density decreases such that hadrons are formed. After this point, the hadrons continue interacting, both inelastically and elastically, until the system is so sparse that there are no further hadronic collisions. The point at which inelastic collisions cease is referred to as chemical freeze-out, and the second stage when elastic collisions can no longer occur is the kinetic freeze-out. After freeze-out the hadrons free stream into the detectors for measurement and identification. Thus, the experimental results for different observables can be linked to the freeze-out stages in the evolution of the system.

Chemical freeze-out parameters are typically obtained by treatment of the particle yields or fluctuations in a thermal model \cite{Acharya2018, Alba2014, Andronic2009}. Thermal fits of particle yields can be used to determine the temperature, baryonic chemical potential, and volume at freeze-out, (${T_f, \mu_{B,f}, V_f}$). If instead, the ratios of the yields are used, then the volume dependence is eliminated. In addition, the fluctuations of conserved charges can be used to determine the freeze-out parameters by comparing experimental results for the particle fluctuations to a thermal model. The Hadron Resonance Gas (HRG) Model has been used to determine the chemical freeze-out conditions in this way \cite{Alba2014}. One study performed an analysis regarding the sensitivity of the freeze-out temperature to the different methods of freeze-out analyses \cite{Alba2015}. It was shown that for many particle species the fluctuations provide a better thermometer than the yields, especially in the case of the strange mesons. This study seeks to characterize the chemical freeze-out during heavy-ion collisions by utilizing the HRG Model to calculate fluctuations of conserved charges \cite{Bellwied2019}.

\section{Methodology}
The chemical freeze-out parameters are determined by using the Hadron Resonance Gas Model. This model describes an interacting gas of hadrons by a system of non-interacting hadrons and their resonant states. Within the Grand Canonical Ensemble, the pressure of such a hadron gas is defined as:

$$\frac{P}{T^4} = \frac{1}{VT^3} \sum_i \, ln \,\! Z_i (T, V, \vec{\mu}),$$

$$\ln \,\! Z^{M/B}_i  = \mp \frac{V \,\! d_i}{(2\pi)^3} \int \!\! d^3 \! k \, \ln \! \left( 1  \mp \, exp \left[ - \left(\epsilon_i - \mu_a X_a^i \right)/T \right] \right),$$
where, the index i runs over all the particles included in the HRG model from the Particle Data Group listing, the energy $\epsilon_i = \sqrt{k^2 + m^2_i}$, conserved charges $\vec{X_i} = \left( B_i, S_i, Q_i \right)$, degeneracy $d_i$, mass $m_i$, and volume $V$.
In this model, the susceptibilities are the derivatives of the pressure with respect to the different chemical potentials associated with the conserved charges in heavy-ion collisions.

$$\chi^{BSQ}_{ijk} = \frac{\partial^{i+j+k} \left( P/T^4 \right)}{\partial \left(\mu_B / T \right)^i \, \partial\left(\mu_S / T \right)^j \, \partial \left(\mu_Q / T \right)^k}$$
The susceptibilities are defined as the various moments of the distribution of the net-charge of interest.

\begin{center}
mean: M = $\chi_1   \qquad \qquad$  variance: $\sigma^2 = \chi_2 $\\ 

skewness: S = $\chi_3/ \left(\chi_2 \right)^{3/2}   \qquad $  kurtosis: ${\kappa = \chi_4/\left(\chi_2 \right)^2}$
\end{center}
These can be directly related to the experimental moments of the net-particle distributions by utilizing the ratios of these quantities:
\\
\begin{center}

M/$\sigma^2$ = $\chi_1 / \chi_2  \qquad \qquad$ S$\sigma$ = $\chi_3 / \chi_2$ \\

$\kappa \sigma^2 = \chi_4 / \chi_2 \qquad \quad  S\sigma^3/M = \chi_3 / \chi_1$ \\

\end{center}

We calculate the ratio $\chi_1/\chi_2$ for the net-kaons ($K^+$ - $K^-$) and compare to the experimental results from the STAR collaboration. In order to directly compare to the experimental data, the HRG model must also include the acceptance cuts on rapidity, y, and transverse momentum, $p_T,$ which match the ones from the experiment. Thus, the susceptibilities for the net-kaon fluctuations are given by:

$$\chi_n^{net-K} = \sum_i^{N_{HRG}} \frac{(Pr_{i \rightarrow net-K})^n}{T^{3-(n-1)}}  \frac{S_i^{1-n}d_i}{4\pi^2} \frac{\partial^{n-1}}{\partial\mu_S^{n-1}} \Bigg\{ \int_{-0.5}^{0.5} d\textit{y} \int_{0.2}^{1.6} d\textit{p}_T \times  $$


$$ \frac{p_T \sqrt {p_T^2 + m_i^2} Cosh[y]}{(-1)^{B_{i}+1} + exp((Cosh[y]\sqrt{p_T^2 + m_i^2} - (B_i\mu_B + S_i\mu_S + Q_i\mu_Q))/T) } \Bigg\} $$ \\
In order to uniquely determine each set of chemical freeze-out parameters for the different collision energies, another quantity from the phase diagram is required. Our first choice was to utilize the higher order moments of the net-kaon distribution, but the error on these quantities is too large to provide a precise determination of the freeze-out parameters. However, the isentropic trajectories from Lattice QCD provide that relationship between T and $\mu_B$ since they show the path of the HIC system through the phase diagram in the absence of dissipation \cite{Guenther2017}. The isentropic lines are established by requiring that the entropy per baryon number ($S/N_B$) is conserved along this path. 

\section{Results} 
We calculate the mean-over-variance, $\chi_1/\chi_2$, of the net-kaons along the Lattice QCD isentropes. By finding the overlap with the experimental values for the same quantity, we determine the freeze-out conditions for the kaons for each collision energy.

\begin{figure}[ht!]
    \centering
    \includegraphics[width=\linewidth]{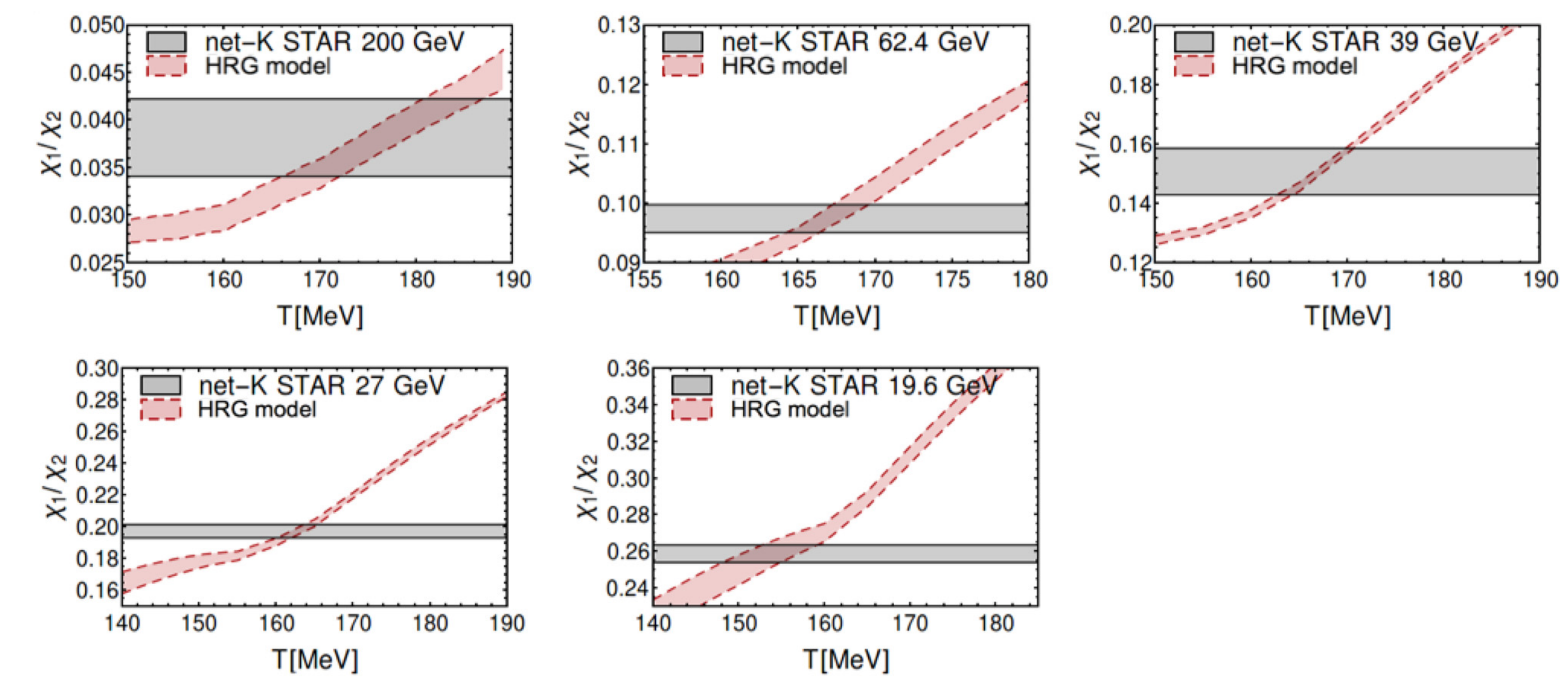}
    \caption{Results for $\chi_1^K/\chi_2^K$ calculated in the HRG model along the lattice QCD isentropic trajectories (pink, dashed band) compared to
${(M/\sigma^2)}^K$ data from [27] (gray, full band) across the Beam Energy Scan at STAR}
    \label{fig:kaon}
\end{figure}

\begin{figure}[ht!]
    \centering
    \includegraphics[width=9.5cm, height=6cm]{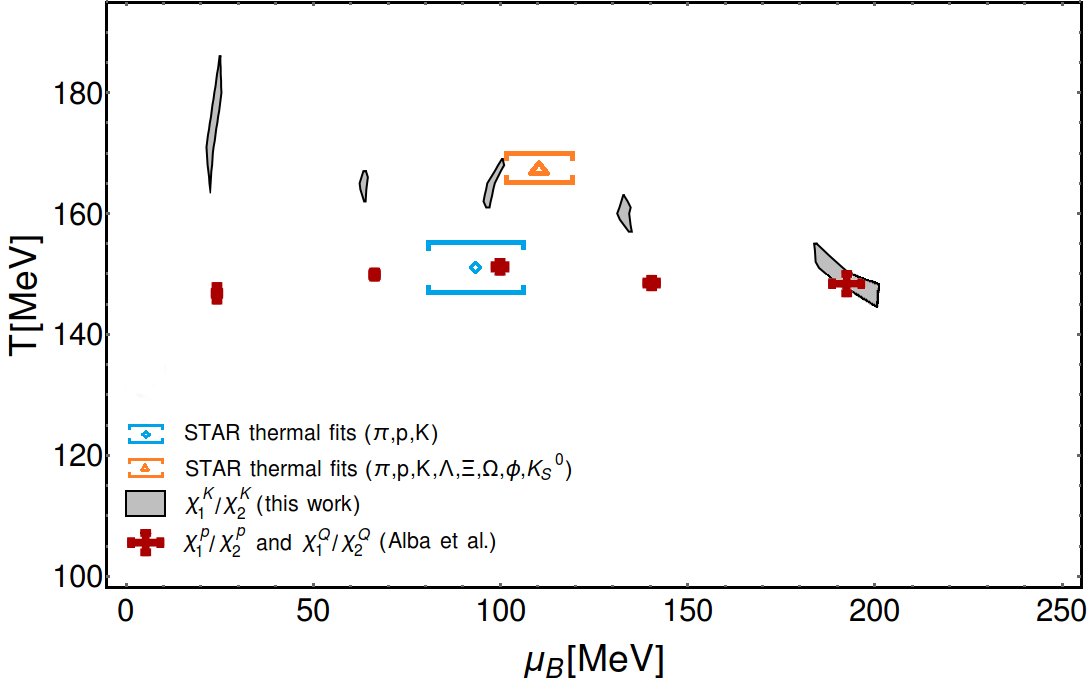}
    \caption{Freeze-out parameters across the highest five energies of the Beam Energy Scan. The red points were obtained from the combined fit of $\chi_1^p/\chi_2^p$ and $\chi_1^Q/\chi_2^Q$, while the gray bands are obtained from the fit of $\chi_1^K/\chi_2^K$ in this work. Also shown are the freeze-out parameters obtained by the STAR collaboration at $\sqrt{s}$ = 39 GeV from thermal fits to all measured ground-state yields (orange triangle) and only to protons, pions, and kaons (blue diamond-shaped symbol).}
    \label{fig:PD}
\end{figure}

\noindent Figure \ref{fig:kaon} shows $\chi_1^K/\chi_2^K$ as a function of the temperature for a range of collision energies of the Beam Energy Scan (BES). In this figure, the red bands are the susceptibilities in the HRG model calculated along the isentropes, and the gray bands correspond to the experimental values with error bars included. For the highest collision energy, we find a freeze-out temperature for net-kaons in the range of T $\approx$ 163-185 MeV. Figure \ref{fig:PD} shows the freeze-out conditions determined from this analysis in the QCD phase diagram as compared to the freeze-out conditions for light hadrons and thermal fits from the STAR collaboration \cite{Adamczyk2017}. The freeze-out conditions for the light hadrons were determined by performing a combined fit of the net-proton and net-electric charge fluctuations. This plot shows there is a separation of about 10-15 MeV between the freeze-out temperature for the kaons and the light hadrons at the highest collision energy.

\begin{figure}[ht!]
    \centering
    \includegraphics[width=10cm, height=6cm]{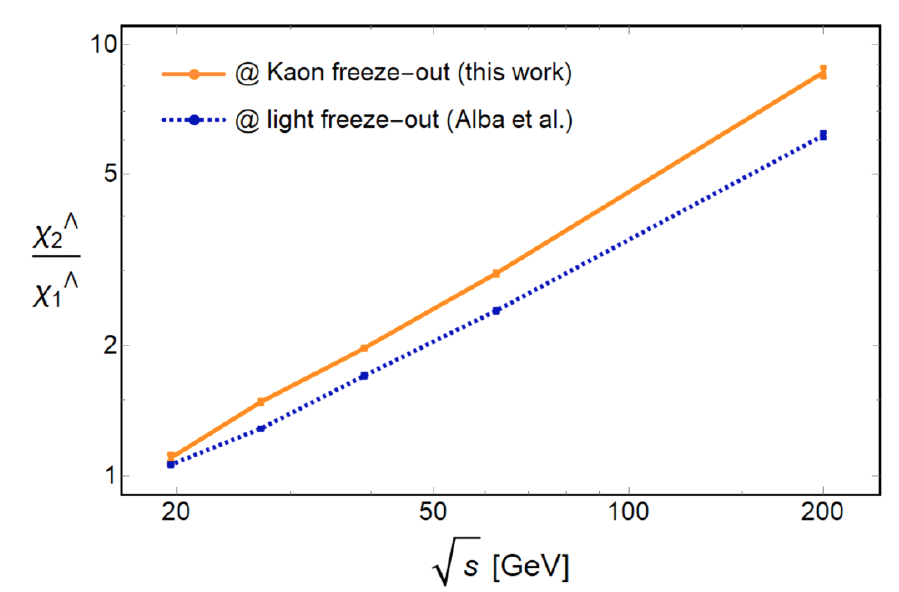}
    \caption{$\chi_2^\Lambda/\chi_1^\Lambda$ as a function of $\sqrt{s}$. The orange points joined by a full line are calculated at the values of $T_f$ and $\mu_B,f$ extracted from the fit of $\chi_1^K/\chi_2^K$, while the blue points joined by a dashed line are calculated at the values of $T_f$ and $\mu_{B,f}$ extracted from the combined fit of $\chi_1^p/\chi_2^p$ and $\chi_1^Q/\chi_2^Q$.}
    \label{fig:lambda}
\end{figure}

\noindent In light of the recent experimental data for the $\Lambda$ fluctuations, we show predictions for the mean-over-variance in figure \ref{fig:lambda} for both freeze-out conditions. The dotted curve shows the $\Lambda$ fluctuations in the case of freeze-out with the light hadrons, and the solid curve shows the fluctuations at a chemical freeze-out with the kaons. Comparison of the experimental data with these results will help to shed light on the behavior of this strange baryon during freeze-out.

\section{Conclusions}
We have presented results for the freeze-out parameters of net-kaons for a range of collision energies of the Beam Energy Scan. We found that the experimental results for the kaons cannot be reproduced by utilizing the freeze-out parameters for the light hadrons, which were determined by the combined fit of net-proton and net-electric charge. We see a separation between the freeze-out temperature for light hadrons and kaons for the highest collision energies of the BES. We provide predictions for the net-$\Lambda$ fluctuations which can be compared to the experimental results coming from the STAR collaboration.

\section{Acknowledgments}
The author would like to acknowledge collaborators on this work: Claudia Ratti, Rene Bellwied, Jaki Noronha-Hostler, Paolo Parotto, and Israel Portillo-Vazquez. This material is based upon work supported by the National
Science Foundation under Grant No. PHY-1654219 and by the U.S. Department of Energy, Office of Science, Office of Nuclear Physics, within the framework of the Beam Energy Scan Theory (BEST) Topical Collaboration. We also acknowledge the support from the Center of Advanced Computing and Data Systems at the University of Houston. 

\end{document}